\documentclass[10pt]{article}
\usepackage{amssymb,amsmath,fullpage,hyperref}

\begin{document}

\begin{figure}
\begin{flushright}
TIFR/TH/11-22
\end{flushright}
\end{figure}

\title{Vacuum statistics and parameter tuning\\
       for F-term supersymmetry breaking}
\author{Zheng Sun\\
        \normalsize\textit{Department of Theoretical Physics, Tata Institute of Fundamental Research,}\\
        \normalsize\textit{1 Homi Bhabha Road, Mumbai 400005, India.}\\
        \normalsize\textit{E-mail:} \texttt{zsun@theory.tifr.res.in}}
\date{}
\maketitle

\begin{abstract}
We carry out a model-independent EFT method study on the vacuum statistics of general F-term SUSY breaking models. Assuming a smooth distribution of Lagrangian parameters, SUSY breaking vacua are rare in global SUSY models with a canonical K\"ahler potential, and have a peaked distribution near the cut-off of the SUSY breaking scale in both global SUSY and SUGRA models with a general K\"ahler potential. After including different mass scales in the Lagrangian, we compare the total number of SUSY and non-SUSY vacua and estimate quantitatively the rareness of SUSY breaking. The EFT method provides a general view to the amount of parameter tuning needed for a metastable SUSY breaking vacuum. The tuning also indicates the importance of R-symmetries in SUSY breaking even for metastable SUSY breaking.
\end{abstract}

\section{Introduction}

Supersymmetry (SUSY) has been proposed for many years to solve several puzzles of the Standard Model. To have a physically acceptable model, SUSY needs to be broken dynamically in a hidden sector \cite{Martin:1997ns, Poppitz:1998vd, Intriligator:2006dd, Intriligator:2007cp, Dine:2010cv}. Then the mass splitting is transferred to the Standard Model sector by different mediation methods \cite{Giudice:1998bp, Meade:2008wd, Dine:2009gy, Kitano:2010fa}. The SUSY breaking scale can be anything from a few hundred GeV to the planck scale, depending on models. To solve the hierarchy problem and gauge coupling unification, and to provide testable predictions for current and near-future experiments, the most interesting models have SUSY breaking at around or not much higher than a TeV.

In the view of the landscape, vast number of models and vacuum states can be constructed from string theory or some other underlying theory \cite{Bousso:2000xa, Kachru:2003aw, Susskind:2003kw, Grana:2005jc},. Each vacuum has its distinct values of physical observables, e.g.\ the SUSY breaking scale, the cosmological constant, etc. If there are enough amount of vacua with their observables distributed around the experimental data, one may believe it is very possible that at least one vacuum has correct values for all observables, and our universe may live on this vacuum through anthropic selection or some evolution processe of the universe. To estimate such possibility, it is important to know the vacuum distribution respecting to these observables. One may study a set of models build from some underlying theory, e.g.\ type IIB flux compactification, and hope they have vacuum statistics similar to the landscape of the ultimate theory. Many results have been reached in this way \cite{Kumar:2006tn, Douglas:2006es, Denef:2007pq}. On the other hand, a simple effective field theory (EFT) with some assumption of parameter distributions known from microscopic theories, is useful to demonstrate as good results as more complicated microscopic theories can achieve \cite{Denef:2004cf, Dine:2005yq}.

We focus on $D = 4$, $N = 1$ SUSY models which are favoured by most phenomenology studies. SUSY can be broken by F-terms or D-terms. The D-term from $\operatorname{U}(1)$ vector fields, i.e.\ the Fayet-Iliopoulos term, is inconsistent with quantum gravity \cite{Komargodski:2009pc, Dienes:2009td}. And non-Abelian D-terms in general can not exist without F-terms of a comparable scale, even for metastable vacua \cite{Dumitrescu:2010ca}. So it is reasonable to concentrate on F-term SUSY breaking. The scale of SUSY breaking can be measured by the magnitude of the F-term field strength
\begin{equation}
F = \sqrt{\sum_i \lvert F_i \rvert^2} , \quad
F_i = \partial_{z_i} W.
\end{equation}
where $W$ is the superpotential for chiral fields $\{z_i\}$. Note that $F$ has mass dimension $2$. For supergravity (SUGRA) partial derivatives are replaced by covariant derivatives.

In section 2, we prescribe the working procedure of the low energy EFT method in a model-independent manner. In section 3, taking the superpotential and the K\"ahler potential to be the most general form, we summarize the vacuum statistics in both global SUSY and SUGRA cases. Assuming a smooth distribution of Lagrangian parameters, we see that for global SUSY models with a canonical K\"ahler potential, SUSY breaking needs a lot of fine-tuning unless R-symmetries are imposed. In models of global SUSY with a general K\"ahler potential and SUGRA with an either canonical or general K\"ahler potential, the number of SUSY breaking vacua below a cutoff $F < F_0$ is always proportional to $F_0^6$, which indicates that the vacuum distribution is always peaked at the highest SUSY breaking scale near the cut-off. The distribution respecting to the cosmological constant is flat in the SUGRA case. These results are consistent with studies in previous literatures \cite{Denef:2004cf, Dine:2005yq, Giudice:2006sn, Sun:2008nh}.

In the string theory landscape studies, one often assumes that there is only one fundamental scale and looks for vacua with low energy SUSY. Viewing our set-up as a more general low energy EFT, it is more suitable to consider that different parts of the Lagrangian may be generated from different dynamics at different energy scales. The presumed mass hierarchy may alter the vacuum statistics. In section 4, we scale the superpotential and K\"ahler potential coefficients to dimensionless by $M_S$ and $M_K$ in addition to the Planck mass $M_P$ which enters the SUGRA formula. We find that the vacuum distribution respecting to the SUSY breaking scale and the cosmological constant is the same as models with only one fundamental scale. The mass hierarchy gives an overall factor to the number of vacua. Comparing to the number of SUSY vacua which is obtained using the same EFT method, the overall factor indicates that non-SUSY vacua are rare by the factor of $M_S^8 / M_K^8$.

In low energy phenomenology studies, rather than having vacuum distributions, It is more important to know the parameter space of models where observables agree with experimental data. The tuning to the allowed parameter region could come from anthropic selection, dynamical process at high energy, cosmological evolution or just fitting the experimental data. For the SUSY breaking sector, the first step is to estimate the size of the parameter space where a metastable SUSY breaking vacuum exists. Our EFT method provides a general view to the amount of tuning, which can be obtained from measuring the constrained parameter region in the previous statistics procedure. In section 5, we see that in the SUGRA case the superpotential gets a large expectation value compared to the SUSY breaking scale, and has to be fine-tuned to get the correct cosmological constant. This is a general feature of gravity mediation. A more common tuning of order $M_S^4 / M_K^4$ in both global SUSY and SUGRA cases comes from the metastability condition. Parameter region of such a size generally exists for models building. The amount of tuning can be viewed as the allowed amount of R-symmetry breaking in models with approximate R-symmetries. This indicates the importance of R-symmetries in SUSY breaking even for metastable vacua, as suggested before \cite{Intriligator:2007py, Abe:2007ax}. We make final comment on the possible extension to multi-field cases.

\section{The general method}

In F-term SUSY breaking models, we have a set of chiral fields $\{z_i | i = 1, \dotsc, d\}$. The superpotential $W$ is a holomorphic function of $\{z_i\}$. We also have a K\"ahler potential $K$ which is a real function of $\{z_i\}$ and $\{\bar{z}_i\}$. They must be smooth at the vacuum $\{z_{i(0)}\}$, so we can expand them near the vacuum in Taylor series. The most general form is
\begin{gather}
W = \sum_{n_i} \frac{1}{n_1! \dotsm n_d!} a_{n_1 \dotsm n_d} (z_1 - z_{1(0)})^{n_1} \dotsm (z_d - z_{d(0)})^{n_d} \ ,\\
K = \sum_{n_i, m_i} \frac{1}{n_1! m_1! \dotsm n_d! m_d!} c_{n_1 m_1 \dotsm n_d m_d} (z_1 - z_{1(0)})^{n_1} (\bar{z}_1 - \bar{z}_{1(0)})^{m_1} \dotsm (z_d - z_{d(0)})^{n_d} (\bar{z}_d - \bar{z}_{d(0)})^{m_d} \ .
\end{gather}
For general consideration, we include all non-renormalizable terms. Summing indices $n$'s and $m$'s go from $0$ to $\infty$, and $i$'s and $j$'s go from $1$ to $d$. To keep $K$ real, the coefficients satisfy
\begin{equation}
c_{n_1 m_1 \dotsm n_d m_d} = c_{m_1 n_1 \dotsm m_d n_d}^* \ .
\end{equation}
We use stars for complex conjugate of coefficients and bars for complex conjugate of other variables.

For global SUSY, we have the scalar potential
\begin{equation}
V = \sum_{i,j} K_{\bar{i} j} \partial_{\bar{z}_i} \bar{W} \partial_{z_j} W
\end{equation}
where $K_{\bar{i} j}$ is the K\"ahler metric which is the inverse of the second derivative matrix of $K$:
\begin{equation} \label{eq:2.0-01}
K_{\bar{i} j} K^{\bar{i} j'} = \delta_j^{j'}, \quad
K^{\bar{i} j} = \partial_{\bar{z}_i} \partial_{z_j} K \ .
\end{equation}
For SUGRA, we have the scalar potential
\begin{equation}
V = e^K (\sum_{i,j} K_{\bar{i} j} D_{\bar{z}_i} \bar{W} D_{z_j} W - 3 \bar{W} W)
\end{equation}
where the K\"ahler metric has the same form as \eqref{eq:2.0-01}, and partial derivatives in the field strength are replaced by SUGRA covariant derivatives:
\begin{equation}
F_i = D_{z_j} W
    = \partial_{z_j} W + W \partial_{z_j} K , \quad
D_{\bar{z}_j} \bar{W} = (D_{z_j} W)^* \ .
\end{equation}
Metastable vacua are found by searching local minima of the scalar potential $V$. Right now all mass scales in Lagrangian parameters are set to $1$, i.e.\ there is presumably only one scale, and we are searching for vacua with $F \ll 1$. We will include different mass scales in later sections.

\subsection{One-field approximation}

The general multi-field model is quite complicated. If there is a field whose mass is much lighter than other fields, one can integrate out heavy fields at a low energy scale, or just freeze their values at leading order. Then one can have an effective one-field theory which is much easier to study. Fortunately this is the case in our consideration. In global SUSY with a canonical K\"ahler potential, there is always a pseudomodulus direction \cite{Ray:2006wk, Komargodski:2009jf}. Its scalar component has zero tree level mass, its fermion component is the massless goldstino, and its auxiliary component is the F-term field strength of SUSY breaking. The pseudomodulus may get non-zero mass from loop corrections, or equivalently, from non-minimal corrections of the K\"ahler potential \cite{Benjamin:2010vm}. The mass is suppressed by the ratio of the SUSY breaking scale to a large mass scale which appears in non-minimal corrections of the K\"ahler potential. The pseudomodulus may also get mass from SUGRA interactions. And the mass is suppressed by the ratio of the SUSY breaking scale to the Planck mass. Such suppression can be seen in later sections.

Other than the pseudomodulus, there is no mechanism to introduce a small mass. The natural value of field masses should be the same order as of Lagrangian parameters, i.e.\ of order $1$. It is possible by tuning parameters to have more than one light fields simultaneously. But such tuning greatly decreases the size of the allowed parameter region \cite{Denef:2004cf}. For this reason, vacua with multiple light fields only contribute a small portion of the landscape unless there is enhancement from symmetries.

Now we can write our effective models of the light field $z$.
\begin{gather} \label{eq:2.1-01}
W = \sum_n \frac{1}{n!} a_n (z - z_0)^n , \quad
K = \sum_{n,m} \frac{1}{n! m!} c_{nm} (\bar{z} - \bar{z_0})^n (z-z_0)^m , \quad
c_{nm} = c_{mn}^* \ ,\\
V = \frac{1}{\bar{\partial} \partial K} \bar{\partial} \bar{W} \partial W \quad
\text{for SUSY},\\
V = e^K (\frac{1}{\bar{\partial} \partial K} \bar{D} \bar{W} D W - 3 \bar{W} W) \quad
\text{for SUGRA},
\end{gather}
where the $\partial$'s and $D$'s are respecting to $z$ or $\bar{z}$:
\begin{equation}
\partial = \frac{\partial}{\partial z} , \quad
\bar{\partial} = \frac{\partial}{\partial \bar{z}} , \quad
D W = \partial W + W \partial K , \quad
\bar{D} \bar{W} = (D W)^* \ .
\end{equation}

\subsection{Constraints on parameters}

SUSY-breaking vacua of our interest should have the following properties:
\begin{enumerate}
\item Small SUSY breaking scale,
\item Stationarity,
\item Metastability,
\item Small cosmological constant.
\end{enumerate}
Each propertie gives some constraint on Lagrangian parameters. In our one-field effective model, coefficients $a_n$ and $c_{nm}$ can take any values of order $1$. But only some region of the parameter space is allowed to have a vacuum of our interest. These constraints can be written as a set of Dirac delta functions $\delta$'s and Heaviside step functions $\Theta$'s. If the vacuum distribution is dense enough on the parameter space so we can take a continuous approximation, the number of vacua can be counted as an integral:
\begin{equation}
N(\text{vacua}) = \int d \mu(a_n, c_{nm}, z) \prod \delta \text{'s and } \Theta \text{'s} \ .
\end{equation}

The small SUSY breaking scale and small cosmological constant conditions can be expressed as $\Theta(F < F_0)$ and $\Theta(0 < V < \Lambda_0)$ where $F_0$ and $\Lambda_0$ are the cut-offs of our interest. For F-term SUSY breaking in global SUSY models, the vacuum energy is always related to the field strength as $V \sim F^2$. But the observed cosmological constant has a much smaller value. SUGRA coupling or other contributions is needed to get a small $V$. We are not considering those extra effects here. So we just drop off the small cosmological constant condition in the global SUSY case. But we include $\Theta(0 < V < \Lambda_0)$ in the SUGRA case.

The stationarity condition is imposed by requiring the first derivative of $V$ to vanish. This can be expressed by the delta function
\begin{equation} \label{eq:2.2-01}
\delta(V') = \delta(\partial V) \delta(\bar{\partial} V) \lvert \det V'' \rvert
\end{equation}
where $V''$ is the mass matrix
\begin{equation} \label{eq:2.2-02}
V''=
\begin{pmatrix}
\bar{\partial} \partial V & \bar{\partial}^2 V\\
\partial^2 V & \bar{\partial} \partial V
\end{pmatrix} \ .
\end{equation}
The form of the delta function gives the correct count of vacua:
\begin{equation} \label{eq:2.2-03}
\int d^2 z \delta(V') = \int d \bar{z} d z \sum_{V'(z_{(0)})=0} \delta^2(z - z_{(0)})
                      = N(\{ z_{(0)} \}) \ .
\end{equation}
In practice, We treat $\partial V$ as a function of some coefficient $a_i$ instead of $z$. The integration of $d^2 a_i$ does not give a simple count of $a_{i(0)}$'s. There is a factor from delta functions:
\begin{equation}
\delta(\partial V) \delta(\bar{\partial} V) = \frac{1}{\lvert J \rvert} \delta^2(a_i - a_{i(0)}) , \quad
J = \frac{\partial (\partial V, \bar{\partial} V)}{\partial (a_i^*, a_i)}
\end{equation}
as well as the factor $\lvert \det V'' \rvert$ from \eqref{eq:2.2-01}. So we have
\begin{equation} \label{eq:2.2-04}
\delta(V') = \frac{\lvert \det V'' \rvert}{\lvert J \rvert} \delta^2(a_i - a_{i(0)}) \ .
\end{equation}

The metastability condition is imposed by requiring the mass matrix $V''$ to be positive definite. In the one field case here, $V''$ is just a $2 \times 2$ Hermitian matrix as \eqref{eq:2.2-02}. So the condition $V''>0$ is equivalent to
\begin{equation}
\Theta(V'' > 0) = \Theta(\bar{\partial} \partial V > 0 , \det V'' > 0) \ .
\end{equation}

With these $\delta$ and $\Theta$ functions, we can write down the total number of vacua which satisfies all the conditions,
\begin{gather}
N(F < F_0) = \int d \mu \Theta(F < F_0) \delta(V') \Theta(V'' > 0) \quad
\text{for SUSY},\\
N(F < F_0, 0 <  V  < \Lambda_0) = \int d \mu \Theta(F < F_0) \delta(V') \Theta(V'' > 0) \Theta(0 < V < \Lambda_0) \quad
\text{for SUGRA}.
\end{gather}
As we are to see in the next section, these conditions give constraints on $\{a_0, \dotsc, a_3\}$ if we treat these coefficients as variables of $\delta$ and $\Theta$ functions. The allowed values of $\{a_0, \dotsc, a_3\}$ are reduced to small intervals. If the distribution of $\{a_0, \dotsc, a_3\}$ is not singular near the allowed region, one can use a uniform distribution to approximate the smooth distribution within such small intervals. Thus the integration of $\{a_0, \dotsc, a_3\}$ picks up a factor proportional to the size of the integration region, which is related to $F_0$ and $\Lambda_0$. Other parameters are not constrained\footnote{One can also leave $a_n$'s unconstrained and tune $c_{nm}$ to fit the vacuum conditions, or spread the constraint in more parameters. The difference is just the choice of variables and does not alter the vacuum distribution result.}. The integration of them just gives a factor which only depends on the form of $d \mu(a_n, c_{nm}, z)$. To know the detailed form of $d \mu$ one needs to study the microscopic theory which produces the landscape. Here we are only interested in the relative distribution respecting to different energy scales. Such a overall factor is not included in this work.

\section{Models and vacuum distributions}

We are to apply the method to the general SUSY and SUGRA models. The effective one-field superpotential and K\"ahler potential are given in \eqref{eq:2.1-01}. From now on we set the vacuum at $z=0$ by a shift, so we have
\begin{equation}
W = \sum_n \frac{1}{n!} a_n z^n , \quad
K = \sum_{n,m} \frac{1}{n! m!} c_{nm} \bar{z}^n z^m \ .
\end{equation}
At the vcuum, the expectation values for their derivatives are:
\begin{equation}
\partial^n W \rvert_{z=0} = a_{n} , \quad
\bar{\partial}^n \partial^m K \rvert_{z=0} = c_{nm} \ .
\end{equation}
We omit $z=0$ in the following notes. Lagrangian quantities are always evaluated at the vacuum.

In the following models, The F-term is\footnote{As we show in later sections, in the SUGRA case, the covariant part of $D W$ vanishes at the vacuum after applying a K\"ahler transformation. So we have $F = \lvert a_1 \rvert$ for both SUSY and SUGRA cases.}
\begin{equation}
F = \lvert \partial W \rvert
  = \lvert a_1 \rvert \ .
\end{equation}
So the small SUSY breaking scale condition always gives
\begin{equation} \label{eq:3.0-01}
\int d^2 a_1 \Theta(F < F_0) \sim \int_0^{F_0} F d F \ .
\end{equation}
We have extracted $d^2 a_1$ from $d \mu$ here since in the small integration region $F \ll 1$, the smooth distribution is approximated with a uniform distribution.

\subsection{SUSY with a canonical K\"ahler potential}

With a canonical K\"ahler potential $K = \bar{z} z$, the scalar potential is simply
\begin{equation}
V = \bar{\partial} \bar{W} \partial W \ .
\end{equation}
And its derivatives of our interest are:
\begin{align}
\partial V &= \bar{\partial} \bar{W} \partial^2 W
            = a_1^* a_2 \ ,\\
\partial^2 V &=  \bar{\partial} \bar{W} \partial^3 W
              = a_1^* a_3 \ ,\\
\bar{\partial} \partial V &= \bar{\partial}^2 \bar{W} \partial^2 W
                           = a_2^* a_2 \ .
\end{align}

Since we are looking for SUSY breaking vacua, we have $a_1 \ne 0$. The stationarity condition requires $a_2 = 0$. Then the diagonal element of the mass matrix $\bar{\partial} \partial V$ vanishes. The off-diagonal element $\partial^2 V$ also needs to vanish to avoid a tachyonic mass eigenvalue. This requires $a_3 = 0$. Now we have a zero mass matrix. If non-renormalizable terms are allowed in $W$, whether the vacuum is metastable is still unknown. To insure that there is no high order instability, one can prove that all coefficients in $W$ except $a_1$ must vanish \cite{Sun:2008nh}. Metastable SUSY breaking vacua can only be achieved from a linear superpotential. This is just a manifestation of the fact that a SUSY breaking vacuum from any O'Raifeartaigh model with a canonical K\"ahler potential always has a flat pseudomodulus direction \cite{Ray:2006wk, Komargodski:2009jf}. In our one-field model the only field $z$ is the pseudomodulus. The integration intervals for $a_n$'s are infinitesimally small for $n>1$, which gives a series of infinitesimally small factors to the total number of vacua. For comparison, SUSY preserving vacua have $a_1=0$ which satisfies stationarity and metastability conditions automatically. There is no constraint except $a_1=0$.

It is well known that SUSY breaking in a global minimum is rare to occur in generic models unless one introduces R-symmetries \cite{Nelson:1993nf}. Here the result indicates that metastable SUSY breaking is also rare in generic models with a canonical K\"ahler potential. If there is an R-symmetry, usually the pseudomodulus $z$ have R-charge $2$. To keep $W$ having R-charge $2$, all $a_n$'s except $a_1$ are suppressed by the R-symmetry, rather than by coincidence or fine-tuning in the case without R-symmetries. This suggests the importance of R-symmetries for SUSY breaking even in a metastable vacuum.

\subsection{SUSY with a general K\"ahler potential}

The scalar potential is
\begin{equation}
V = \frac{1}{\bar{\partial} \partial K} \bar{\partial} \bar{W} \partial W \ .
\end{equation}
And its derivatives of our interest are:
\begin{align}
\begin{split}
\partial V &= \frac{1}{\bar{\partial} \partial K}
              \bar{\partial} \bar{W} \partial^2 W
              - \frac{\bar{\partial} \partial^2 K}{(\bar{\partial} \partial K)^2}
                \bar{\partial} \bar{W} \partial W\\
           &= c_{11}^{-1} a_1^* a_2 - c_{11}^{-2} c_{12} a_1^* a_1 \ ,
\end{split}\\
\begin{split}
\partial^2 V &= \frac{1}{\bar{\partial} \partial K}
                \bar{\partial} \bar{W} \partial^3 W
                - \frac{2 \bar{\partial} \partial^2 K}{(\bar{\partial} \partial K)^2}
                  \bar{\partial} \bar{W} \partial^2 W
                + (\frac{2 (\bar{\partial} \partial^2 K)^2}{(\bar{\partial} \partial K)^3}
                   - \frac{\bar{\partial} \partial^3 K}{(\bar{\partial} \partial K)^2})
                  \bar{\partial} \bar{W} \partial W\\
             &= c_{11}^{-1} a_1^* a_3 - 2 c_{11}^{-2} c_{12} a_1^* a_2 + (2 c_{11}^{-3} c_{12}^2 - c_{11}^{-2} c_{13}) a_1^* a_1 \ ,
\end{split}\\
\begin{split}
\bar{\partial} \partial V &= \frac{1}{\bar{\partial} \partial K}
                             \bar{\partial}^2 \bar{W} \partial^2 W
                             - \frac{\bar{\partial}^2 \partial K}{(\bar{\partial} \partial K)^2}
                               \bar{\partial} \bar{W} \partial^2 W
                             - \frac{\bar{\partial} \partial^2 K}{(\bar{\partial} \partial K)^2}
                               \bar{\partial}^2 \bar{W} \partial W
                             +\\
                &\hphantom{=}
                             + (\frac{2 \bar{\partial}^2 \partial K \bar{\partial} \partial^2 K}{(\bar{\partial} \partial K)^3}
                                - \frac{\bar{\partial}^2 \partial^2 K}{(\bar{\partial} \partial K)^2})
                               \bar{\partial} \bar{W} \partial W\\
                          &= c_{11}^{-1} a_2^* a_2 - c_{11}^{-2} c_{12}^* a_1^* a_2 - c_{11}^{-2} c_{12} a_1 a_2^* + (2 c_{11}^{-3} c_{12}^* c_{12} - c_{11}^{-2} c_{22}) a_1^* a_1 \ .
\end{split}
\end{align}
Usually the K\"ahler potential is written as a canonical form plus corrections, i.e.\ we set $c_{11} = 1$ by a field redefinition. These quantities can be simplified to
\begin{align}
\partial V &= a_1^* a_2 - c_{12} a_1^* a_1 \ ,\\
\partial^2 V &= a_1^* a_3 - 2 c_{12} a_1^* a_2 + (2 c_{12}^2 - c_{13}) a_1^* a_1 \ ,\\
\bar{\partial} \partial V &= a_2^* a_2 - c_{12}^* a_1^* a_2 - c_{12} a_1 a_2^* + (2 c_{12}^* c_{12} - c_{22}) a_1^* a_1 \ .
\end{align}

We have $a_1 \ne 0$. The stationarity condition requires $a_2 = c_{12} a_1$. The typical value for $c_{nm}$ is of order $1$ since they are not going to be constrained in our treatment. So both $a_1$ and $a_2$ are at the SUSY breaking scale $a_1 \sim a_2 \sim F$. Then we know that the diagonal element of the mass matrix $\bar{\partial} \partial V$ is of order $F^2$. For any values of $a_n$ and $c_{nm}$, $a_2^* a_2$ and $2 c_{12}^* c_{12} a_1^* a_1$ are positive-definite. The rest terms in $\bar{\partial} \partial V$ can be positive or negative. So roughly half of the parameter space can make $\bar{\partial} \partial V > 0$. To make $\det V'' > 0$, the off-diagonal element $\partial^2 V$ also needs to be of order $F^2$ or less, which requires $a_3 \lesssim F$.

Now we have
\begin{equation}
a_3 \lesssim a_1
        \sim a_2
        \sim F \ .
\end{equation}
Treating $\partial V$ as a function of $a_2$, we have
\begin{equation}
\lvert J \rvert = \left \lvert \frac{\partial (\partial V, \bar{\partial} V)}{\partial (a_2^*, a_2)} \right \rvert
                = \left \lvert \frac{\partial (\partial V)}{\partial a_2} \right \rvert^2
                = 4 a_1^* a_1
             \sim F^2 \ .
\end{equation}
From the estimation of elements of $V''$, we know
\begin{equation}
\lvert \det V'' \rvert \sim F^4 \ .
\end{equation}
So the stationarity condition gives
\begin{equation} \label{eq:3.2-01}
\int d^2 a_2 \delta(V') = \int d^2 a_2 \frac{\lvert \det V'' \rvert}{\lvert J \rvert} \delta^2 (a_2 - a_{2(0)})
                     \sim \int d^2 a_2 \frac{F^4}{F^2} \delta^2 (a_2 - a_{2(0)})
                     \sim F^2 \ .
\end{equation}
And the metastability condition gives
\begin{equation} \label{eq:3.2-02}
\int d^2 a_3 \Theta(V'' > 0) \sim \int d^2 a_3 \Theta(\det V'' > 0)
                             \sim \int d^2 a_3 \Theta(a_3 \lesssim F)
                             \sim F^2 \ .
\end{equation}
Note the other part of the metastability condition $\Theta( \bar{\partial} \partial V > 0)$ only reduces roughly half of the integration region, thus gives an order $1$ factor which we are not interested in here.

Combining \eqref{eq:3.0-01}\eqref{eq:3.2-01}\eqref{eq:3.2-02} we get the number of SUSY-breaking vacua:
\begin{equation} \label{eq:3.2-03}
N(F < F_0) \sim \int d^6 a_1 \dotsm a_3 \Theta(F < F_0) \delta(V') \Theta(V'' > 0)
           \sim \int_0^{F_0} F \cdot F^2 \cdot F^2 d F
           \sim F_0^6 \ .
\end{equation}
The power of $F_0$ indicates that the distribution is peaked at the highest SUSY breaking scale. Most states live around the cut-off $F \sim F_0$. In reality, the cut-off is where new physics starts to be important, and the low energy EFT becomes not so valid. For example, in the non-SUSY branch of the landscape of type IIB flux compactification \cite{Denef:2004cf, Dine:2005yq} (which is actually studied as a SUGRA EFT which we are to discuss in the next subsection), the cut-off is near the string scale where one has to use the original ten-dimensional theory. The vacuum distribution near the cut-off is unknown in the EFT point of view. The distribution from the EFT method is more trustable at scales much smaller than the cut-off.

\subsection{SUGRA with a general K\"ahler potential}

The scalar potential is
\begin{equation}
V = e^K (\frac{1}{\bar{\partial} \partial K} \bar{D} \bar{W} D W - 3 \bar{W} W)
  = e^{c_{00}} (c_{11}^{-1} (a_1^* + a_0^* c_{01}^*) (a_1 + a_0 c_{01}) - 3 a_0^* a_0) \ .
\end{equation}
These quantities are needed for the convenience of our calculation:
\begin{align}
D W &= \partial W + W \partial K
     = a_1 + a_0 c_{01} \ ,\\
\partial D W &= \partial^2 W + \partial W \partial K + W \partial^2 K
              = a_2 + a_1 c_{01} + a_0 c_{02} \ ,\\
\partial^2 D W &= \partial^3 W + \partial^2 W \partial K + 2 \partial W \partial^2 K + W \partial^3 K
                = a_3 + a_2 c_{01} + 2 a_1 c_{02} + a_0 c_{03} \ ,\\
\bar{\partial} D W &= W \bar{\partial} \partial K \ ,\\
\bar{\partial}^2 D W &= W \bar{\partial}^2 \partial K \ ,\\
\bar{\partial} \partial D W &= \partial W \bar{\partial} \partial K + W \bar{\partial} \partial^2 K \ .
\end{align}
The derivatives of $V$ of our interest are:
\begin{align}
\begin{split}
\partial V &= e^K (\frac{1}{\bar{\partial} \partial K}
                   \bar{D} \bar{W} \partial D W
                   + (\frac{\partial K}{\bar{\partial} \partial K}
                      - \frac{\bar{\partial} \partial^2 K}{(\bar{\partial} \partial K)^2})
                     \bar{D} \bar{W} D W
                   - 2 \bar{W} D W)\\
           &= e^{c_{00}} (c_{11}^{-1}
                          (a_1^* + a_0^* c_{01}^*) (a_2 + a_1 c_{01} + a_0 c_{02}) +\\
 &\hphantom{= e^{c_{00}} (}
                          + (c_{11}^{-1} c_{01} - c_{11}^{-2} c_{12})
                            (a_1^* + a_0^* c_{01}^*) (a_1 + a_0 c_{01}) +\\
 &\hphantom{= e^{c_{00}} (}
                          - 2 a_0^* (a_1 + a_0 c_{01})) \ ,
\end{split}\\
\begin{split}
\partial^2 V &= e^K (\frac{1}{\bar{\partial} \partial K}
                     \bar{D} \bar{W} \partial^2 D W
                     + (\frac{2 \partial K}{\bar{\partial} \partial K}
                        - \frac{2 \bar{\partial} \partial^2 K}{(\bar{\partial} \partial K)^2})
                       \bar{D} \bar{W} \partial D W
                     - \bar{W} \partial D W
                     +\\
   &\hphantom{= e^K (} 
                     + (\frac{(\partial K)^2}{\bar{\partial} \partial K}
                        + \frac{\partial^2 K}{\bar{\partial} \partial K}
                        - \frac{2 \partial K \bar{\partial} \partial^2 K}{(\bar{\partial} \partial K)^2}
                        - \frac{\bar{\partial} \partial^3 K}{(\bar{\partial} \partial K)^2}
                        + \frac{2 (\bar{\partial} \partial^2 K)^2}{(\bar{\partial} \partial K)^3})
                       \bar{D} \bar{W} D W
                     +\\
   &\hphantom{= e^K (}
                     - (\partial K
                        + \frac{\bar{\partial} \partial^2 K}{\bar{\partial} \partial K})
                       \bar{W} D W)\\
             &= e^{c_{00}} (c_{11}^{-1}
                            (a_1^* + a_0^* c_{01}^*) (a_3 + a_2 c_{01} + 2 a_1 c_{02} + a_0 c_{03}) +\\
   &\hphantom{= e^{c_{00}} (}
                            + (2 c_{11}^{-1} c_{01} - 2 c_{11}^{-2} c_{12})
                              (a_1^* + a_0^* c_{01}^*) (a_2 + a_1 c_{01} + a_0 c_{02}) +\\
   &\hphantom{= e^{c_{00}} (}
                            - a_0^* (a_2 + a_1 c_{01} + a_0 c_{02}) +\\
   &\hphantom{= e^{c_{00}} (}
                            + (c_{11}^{-1} c_{01}^2 + c_{11}^{-1} c_{02} - 2 c_{11}^{-2} c_{01} c_{12} - c_{11}^{-2} c_{13} + 2 c_{11}^{-3} c_{12}^2)
                              (a_1^* + a_0^* c_{01}^*) (a_1 + a_0 c_{01}) +\\
   &\hphantom{= e^{c_{00}} (}
                            - (c_{01} + c_{11}^{-1} c_{12})
                              a_0^* (a_1 + a_0 c_{01})) \ ,
\end{split}\\
\begin{split}
\bar{\partial} \partial V &= e^K (\frac{1}{\bar{\partial} \partial K}
                                  \bar{\partial} \bar{D} \bar{W} \partial D W
                                  + (\frac{\bar{\partial} K}{\bar{\partial} \partial K}
                                     - \frac{\bar{\partial}^2 \partial K}{(\bar{\partial} \partial K)^2})
                                    \bar{D} \bar{W} \partial D W
                                  + (\frac{\partial K}{\bar{\partial} \partial K}
                                     - \frac{\bar{\partial} \partial^2 K}{(\bar{\partial} \partial K)^2})
                                    \bar{\partial} \bar{D} \bar{W} D W
                                  +\\
                &\hphantom{= e^K (}
                                  + (\frac{\bar{\partial} K \partial K}{\bar{\partial} \partial K}
                                     - \frac{\bar{\partial} K \bar{\partial} \partial^2 K}{(\bar{\partial} \partial K)^2}
                                     - \frac{\partial K \bar{\partial}^2 \partial K}{(\bar{\partial} \partial K)^2}
                                     - \frac{\bar{\partial}^2 \partial^2 K}{(\bar{\partial} \partial K)^2}
                                     + \frac{2 \bar{\partial}^2 \partial K \bar{\partial} \partial^2 K}{(\bar{\partial} \partial K)^3})
                                     \bar{D} \bar{W} D W
                                  +\\
                &\hphantom{= e^K (}
                                  - 2 \bar{\partial} \partial K
                                    \bar{W} D W)\\
                          &= e^{c_{00}} (c_{11}^{-1}
                                         (a_2^* + a_1^* c_{01}^* + a_0^* c_{02}^*) (a_2 + a_1 c_{01} + a_0 c_{02}) +\\
                &\hphantom{= e^{c_{00}} (}
                                         + (c_{11}^{-1} c_{01}^* - c_{11}^{-2} c_{12}^*)
                                           (a_1^* + a_0^* c_{01}^*) (a_2 + a_1 c_{01} + a_0 c_{02}) +\\
                &\hphantom{= e^{c_{00}} (}
                                         + (c_{11}^{-1} c_{01} - c_{11}^{-2} c_{12})
                                           (a_1 + a_0 c_{01}) (a_2^* + a_1^* c_{01}^* + a_0^* c_{02}^*) +\\
                &\hphantom{= e^{c_{00}} (}
                                         + (c_{11}^{-1} c_{01}^* c_{01} - c_{11}^{-2} c_{01}^* c_{12} - c_{11}^{-2} c_{01} c_{12}^* - c_{11}^{-2} c_{22} + 2 c_{11}^{-3} c_{12}^* c_{12}) \times\\
                &\hphantom{= e^{c_{00}} (+}
                                           \times (a_1^* + a_0^* c_{01}^*) (a_1 + a_0 c_{01}) +\\
                &\hphantom{= e^{c_{00}} (}
                                         - 2 c_{11} a_0^* a_0) \ .
\end{split}
\end{align}
If we apply a K\"ahler transformation
\begin{equation}
W \to e^{-h} W , \quad
K \to K + h^* + h 
\end{equation}
to set $c_{0n} = c_{n0} = 0$, and set the minimal term of $K$ to the canonical form $c_{11} = 1$ by a field redefinition. $V$ and its derivatives can be simplified to
\begin{align} \label{eq:3.3-01}
V &= a_1^* a_1 - 3 a_0^* a_0 \ ,\\
\partial V &= a_1^* a_2 - c_{12} a_1^* a_1 - 2 a_0^* a_1 \ ,\\
\partial^2 V &= a_1^* a_3 - 2 c_{12} a_1^* a_2 + (2 c_{12}^2 - c_{13}) a_1^* a_1 - a_0^* a_2 - c_{12} a_0^* a_1 \ ,\\
\label{eq:3.3-02}
\bar{\partial} \partial V &= a_2^* a_2 - c_{12}^* a_1^* a_2 - c_{12} a_1 a_2^* + (2 c_{12}^* c_{12} - c_{22}) a_1^* a_1 - 2 a_0^* a_0 \ .
\end{align}

We have $D W = a_1$ at the vacuum after the K\"ahler transformation. So we can continue using the same expression \eqref{eq:3.0-01} for the small SUSY breaking scale condition. The small cosmological constant condition requires
\begin{equation} \label{eq:3.3-03}
\lvert a_1 \rvert > \sqrt{3} \lvert a_0 \rvert
                  > \sqrt{\lvert a_1 \rvert^2 - \Lambda_0} \ .
\end{equation}
Since the observed cosmological constant is much smaller than the SUSY breaking scale, i.e.\ $\Lambda_0 \ll F^2$, we have $a_0 \sim a_1 \sim F$. The stationarity condition requires
\begin{equation}
a_2 = c_{12} a_1 + a_0^* \frac{a_1}{a_1^*}
\end{equation}
which set $a_2 \sim F$. Then we know that the diagonal element of the mass matrix $\bar{\partial} \partial V$ is of order $F^2$. Roughly half of the parameter space can make $\bar{\partial} \partial V > 0$. To make $\det V'' > 0$, the off-diagonal element $\partial^2 V$ also needs to be of order $F^2$ or less, which requires $a_3 \lesssim F$.

\begin{figure}
\setlength{\unitlength}{56pt}
\centering
\begin{picture}(4,2)(-1,-1)
\put(0,0){\vector(1,0){2}}
\put(2.05,0){$a_0$}
\put(0,0){\vector(1,1){0.707106781}}
\put(0.757106781,0.707106781){$\frac{1}{\sqrt{3}} F$ (or $\frac{M_P}{\sqrt{3} M_S} F$)}
\put(0.9,0){\line(0,-1){0.9}}
\put(1,0){\line(0,-1){0.9}}
\put(0.65,-0.85){\vector(1,0){0.25}}
\put(0.9,-0.85){\line(1,0){0.1}}
\put(1.25,-0.85){\vector(-1,0){0.25}}
\put(1.3,-0.85){$\frac{\Lambda_0}{2 \sqrt{3} F}$ (or $\frac{\Lambda_0 M_P}{2 \sqrt{3} F M_S^5}$)}
%\put(0,0){\circle{1}}
\qbezier(-1,0)(-1,-0.414213562)(-0.707106781,-0.707106781)
\qbezier(-0.707106781,-0.707106781)(-0.414213562,-1)(0,-1)
\qbezier(0,-1)(0.414213562,-1)(0.707106781,-0.707106781)
\qbezier(0.707106781,-0.707106781)(1,-0.414213562)(1,0)
\qbezier(1,0)(1,0.414213562)(0.707106781,0.707106781)
\qbezier(0.707106781,0.707106781)(0.414213562,1)(0,1)
\qbezier(0,1)(-0.414213562,1)(-0.707106781,0.707106781)
\qbezier(-0.707106781,0.707106781)(-1,0.414213562)(-1,0)
%\put(0,0){\circle{0.9}}
\qbezier(-0.9,0)(-0.9,-0.372792206)(-0.636396103,-0.636396103)
\qbezier(-0.636396103,-0.636396103)(-0.372792206,-0.9)(0,-0.9)
\qbezier(0,-0.9)(0.372792206,-0.9)(0.636396103,-0.636396103)
\qbezier(0.636396103,-0.636396103)(0.9,-0.372792206)(0.9,0)
\qbezier(0.9,0)(0.9,0.372792206)(0.636396103,0.636396103)
\qbezier(0.636396103,0.636396103)(0.372792206,0.9)(0,0.9)
\qbezier(0,0.9)(-0.372792206,0.9)(-0.636396103,0.636396103)
\qbezier(-0.636396103,0.636396103)(-0.9,0.372792206)(-0.9,0)
\end{picture}
\caption{The value of $a_0$ is restricted in the ring-shaped area in the complex plane to satisfy the small cosmological constant condition. The size of the area after including mass scales is also given in parentheses.} \label{fg:1}
\end{figure}
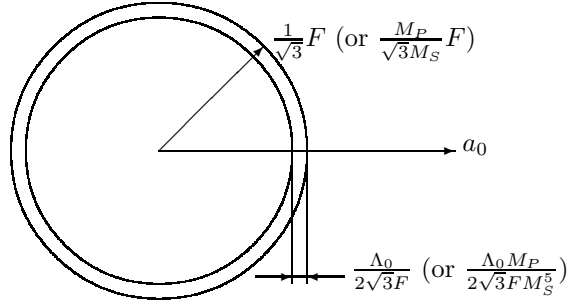

Now we have
\begin{equation}
a_3 \lesssim a_0
    \sim a_1
    \sim a_2
    \sim F
\end{equation}
Following the same procedure for the SUSY case in last subsection, the stationarity condition gives
\begin{equation} \label{eq:3.3-04}
\int d^2 a_2 \delta(V') = \int d^2 a_2 \frac{\lvert \det V'' \rvert}{\lvert J \rvert} \delta^2 (a_2 - a_{2(0)})
                     \sim \int d^2 a_2 \frac{F^4}{F^2} \delta^2 (a_2 - a_{2(0)})
                     \sim F^2 \ .
\end{equation}
And the metastability condition gives
\begin{equation} \label{eq:3.3-05}
\int d^2 a_3 \Theta(V'' > 0) \sim \int d^2 a_3 \Theta(a_3 \lesssim F)
                             \sim F^2 \ .
\end{equation}
From \eqref{eq:3.3-03} we know the integration region of $a_0$ is restricted in
\begin{equation}
\lvert a_0 \rvert \in (\frac{1}{\sqrt{3}} \sqrt{F^2 - \Lambda_0}, \frac{1}{\sqrt{3}} F)
              \approx (\frac{1}{\sqrt{3}} (F - \frac{\Lambda_0}{2 F}), \frac{1}{\sqrt{3}} F)
\end{equation}
as shown in figure \ref{fg:1}. So the small cosmological constant condition gives
\begin{equation} \label{eq:3.3-06}
\int d^2 a_0 \Theta(0 <  V  < \Lambda_0) \sim 2 \pi \frac{F}{\sqrt{3}} \cdot \frac{\Lambda_0}{2 \sqrt{3} F}
                                         \sim \Lambda_0 \ .
\end{equation}

Combining \eqref{eq:3.0-01}\eqref{eq:3.3-04}\eqref{eq:3.3-05}\eqref{eq:3.3-06} we get the number of SUSY-breaking vacua:
\begin{equation} \label{eq:3.3-07}
\begin{split}
N(F < F_0, 0 <  V  < \Lambda_0) &\sim \int d^8 a_0 \dotsm a_3 \Theta(F < F_0) \delta(V') \Theta(V'' > 0) \Theta (0 <  V  < \Lambda_0)\\
                                &\sim \int_0^{F_0} F \cdot F^2 \cdot F^2 \cdot \Lambda_0 d F
                                 \sim F_0^6 \Lambda_0 \ .
\end{split}
\end{equation}
Similarly to the result \eqref{eq:3.2-03} of the global SUSY case, the power of $F_0$ indicates that the distribution is peaked at the highest SUSY breaking scale. Most states live near the cut-off $F \sim F_0$ where the distribution is unknown in the EFT point of view. The distribution is more trustable at scales much smaller than the cut-off. The distribution respecting to the cosmological constant is flat.

Note that if the K\"ahler potential is canonical, then
\begin{align}
V &= a_1^* a_1 - 3 a_0^* a_0 \ ,\\
\partial V &= 2 a_1^* a_2 - 2 a_0^* a_1 \ ,\\
\partial^2 V &= 6 a_1^* a_3 - 2 a_0^* a_2 \ ,\\
\bar{\partial} \partial V &= 4 a_2^* a_2 - 2 a_0^* a_0 \ .
\end{align}
Following the same analysis we still get $a_3 \lesssim a_0 \sim a_1 \sim a_2 \sim F$. The vacuum distribution is the same as \eqref{eq:3.3-07}.

\section{Including mass scales}

In the previous section we set all mass scales to $1$ and look for vacua with $F \ll 1$. This is the case of the non-SUSY branch of the type IIB flux compactification landscape where we set the string scale to $1$ and look for vacua with a small SUSY breaking scale. In phenomenology model building, one usually starts with a low energy theory which already have mass hierarchy, and expects some high energy dynamics could generate such hierarchy. It would be interesting to see how the statistical analysis of the last section applies to such phenomenology-friendly models. Here we assume that there are $3$ different mass scales: $M_S$ appearing in the superpotential $W$, which is related to SUSY dynamics; $M_K$ appearing in non-minimal corrections of the K\"ahler potential $K$, which may, depending on models, come from gauge dynamics, compactified dimensions or integrated-out heavy fields; and the Planck mass $M_P$ appearing in the SUGRA Lagrangian. Note that $W$, $K$, $V$ and $z$ have mass dimensions $3$, $2$, $4$ and $1$ respectively. We can write down the general one-field SUGRA EFT with these mass scales:
\begin{gather}
W = \sum_n \frac{1}{n!} a_n M_S^{3 - n} z^n , \quad
K = \sum_{n,m} \frac{1}{n! m!} c_{nm} M_K^{2 - n - m} \bar{z}^n z^m \ ,\\
V = e^{\frac{K}{M_P^2}} (\frac{1}{\bar{\partial} \partial K} \bar{D} \bar{W} D W - \frac{3 \bar{W} W}{M_P}), \quad
D W = \partial W + \frac{W \partial K}{M_P^2} \ .
\end{gather}
Now all coefficients $a_n$ and $c_{nm}$ are dimensionless. We also redefine $F$ to be dimensionless so the SUSY breaking field strength is $F M_S^2$. With the newly introduced scale $M_S$, $F$ is not necessarily small. We are more interested in vacua with $F$ of order $1$. So the following analysis is a little different than the previous section.

\subsection{Vacuum distributions}

One can do the same calculation as in the last section to get the derivatives of $V$ of our interest. But there is a easier way. We take the expressions \eqref{eq:3.3-01}--\eqref{eq:3.3-02}, do the following replacement:
\begin{equation}
a_n \to a_n M_S^{3 - n} , \quad
c_{nm} \to c_{nm} M_K^{2 - n - m} \ ,
\end{equation}
and insert powers of $M_P$ to where the mass dimension does not match. Then we get
\begin{align} \label{eq:4.1-01}
V &= a_1^* a_1 M_S^4
     - 3 a_0^* a_0 \frac{M_S^6}{M_P^2} \ ,\\
\partial V &= a_1^* a_2 M_S^3
              - c_{12} a_1^* a_1 \frac{M_S^4}{M_K}
              - 2 a_0^* a_1 \frac{M_S^5}{M_P^2} \ ,\\
\partial^2 V &= a_1^* a_3 M_S^2
                - 2 c_{12} a_1^* a_2 \frac{M_S^3}{M_K}
                + (2 c_{12}^2 - c_{13}) a_1^* a_1 \frac{M_S^4}{M_K^2}
                - a_0^* a_2 \frac{M_S^4}{M_P^2}
                - c_{12} a_0^* a_1 \frac{M_S^5}{M_K M_P^2} \ ,\\
\label{eq:4.1-02}
\bar{\partial} \partial V &= a_2^* a_2 M_S^2
                             - (c_{12}^* a_1^* a_2
                             + c_{12} a_1 a_2^*) \frac{M_S^3}{M_K}
                             + (2 c_{12}^* c_{12} - c_{22}) a_1^* a_1 \frac{M_S^4}{M_K^2}
                             - 2 a_0^* a_0 \frac{M_S^6}{M_P^4} \ .
\end{align}

There is one issue we would like to clarify before we continue. The results \eqref{eq:3.3-01}--\eqref{eq:3.3-02} and \eqref{eq:4.1-01}--\eqref{eq:4.1-02} have been simplified by several field redefinitions. In the previous section all mass scales are set to $1$. After field redefinitions, values of Lagrangian parameters change amount of order $1$ and no singularity is introduced to the parameter distribution. So one can continue the analysis using the newly defined fields. This is not so obvious with mass scales included. For the translation $(z-z_0) \to z$, the typical value of $z_0$ is of order $M_S$. After the translation, values of $a_n$'s change amount of order $1$, and values of $c_{nm}$'s change amount of order $M_S / M_K$. The K\"ahler transformation to set $c_{0n} = c_{n0} = 0$ is
\begin{equation}
W \to e^{-\frac{h}{M_P^2}} W , \quad
K \to K + h^* + h , \quad
h =  - \sum_n \frac{1}{n!} c_{0n} M_K^{2 - n} z^n
\end{equation}
where we have included mass scales and the required form of $h$ is given. Giving the condition $M_K < M_P$, values of $a_n$'s change amount of order $M_K^2 / M_P^2$ at the leading order. Finally, the field redefinition to set $c_{11} = 1$ scales all coefficients by factors of order $1$. In summary, values of dimensionless Lagrangian parameters change amount up to order $1$ after all field redefinitions. If we start with a smooth distribution of Lagrangian parameters, the distribution after field redefinitions is also smooth and can be approximated by a uniform distribution. So it is valid to use simplified results \eqref{eq:4.1-01}--\eqref{eq:4.1-02} for distribution analysis.

We still have $F = \lvert a_1 \rvert$ after the K\"ahler transformation. So the small SUSY breaking scale condition gives the same factor as \eqref{eq:3.0-01}. The small cosmological constant condition requires
\begin{equation} \label{eq:4.1-03}
\lvert a_1 \rvert > \frac{\sqrt{3} M_S}{M_P} \lvert a_0 \rvert
                  > \sqrt{\lvert a_1 \rvert^2 - \frac{\Lambda_0}{M_S^4}} \ .
\end{equation}
Since the observed cosmological constant is much smaller than the SUSY breaking scale, i.e.\ $\Lambda_0 \ll F^2 M_S^4$, we have
\begin{equation}
a_0 \sim F \frac{M_P}{M_S} \ .
\end{equation}
The stationarity condition requires
\begin{equation}
a_2 = c_{12} a_1 \frac{M_S}{M_K} + 2 a_0^* \frac{a_1}{a_1^*} \frac{M_S^2}{M_P^2}
 \sim F M_S (\frac{1}{M_K} + \frac{1}{M_P}) \ .
\end{equation}
Then we can estimate the magnitude of the diagonal element of the mass matrix:
\begin{equation}
\bar{\partial} \partial V \sim F^2 M_S^4 (\frac{1}{M_K^2} + \frac{1}{M_P^2}) \ .
\end{equation}
From now on, conditions $M_S \ll M_K$ and $M_S \ll M_P$ are imposed to keep only the lowest order terms in estimations. Although in most realistic models $M_K$ is either much smaller than or at the same order as $M_P$, their magnitude relation is not fixed right now (and is to be studied in later subsections). So we keep both lowest order terms in the result. Roughly half of the parameter space can make $\bar{\partial} \partial V > 0$. To make $\det V'' > 0$, the off-diagonal element $\partial^2 V$ also needs to be of the same order or less. We have
\begin{equation}
\partial^2 V \sim a_3 F M_S^2 + F^2 M_S^4 (\frac{1}{M_K^2} + \frac{1}{M_P^2})
         \lesssim F^2 M_S^4 (\frac{1}{M_K^2} + \frac{1}{M_P^2}) \ ,
\end{equation}
which requires
\begin{equation} \label{eq:4.1-04}
a_3 \lesssim F M_S^2 (\frac{1}{M_K^2} + \frac{1}{M_P^2}) \ .
\end{equation}

The field $z$ has mass dimension $1$, which makes the previous count of vacua \eqref{eq:2.2-03} having mass dimension $-2$. To make a dimensionless count one needs to rewrite the delta function as
\begin{equation}
\delta^2(z - z_{(0)}) \to \delta^2(\frac{1}{M_S} (z - z_{(0)}))
                        = M_S^2 \delta^2(z - z_{(0)}) \ .
\end{equation}
So $\delta(V')$ picks up a factor of $M_S^2$ compared to \eqref{eq:2.2-01} and \eqref{eq:2.2-04}:
\begin{equation}
\delta(V') = M_S^2 \delta(\partial V) \delta(\bar{\partial} V) \lvert \det V'' \rvert
           = M_S^2 \frac{\lvert \det V'' \rvert}{\lvert J \rvert} \delta^2 (a_i - a_{i(0)}) \ .
\end{equation}
Treating $\partial V$ as a function of $a_2$, we have
\begin{equation}
\lvert J \rvert = \left \lvert \frac{\partial (\partial V, \bar{\partial} V)}{\partial (a_2^*, a_2)} \right \rvert
                = \left \lvert \frac{\partial (\partial V)}{\partial a_2} \right \rvert^2
                = 4 a_1^* a_1 M_S^6
             \sim F^2 M_S^6 \ .
\end{equation}
From estimations of elements of $V''$, we have
\begin{equation}
\lvert \det V'' \rvert \sim F^4 M_S^8 (\frac{1}{M_K^4} + \frac{1}{M_P^4}) \ .
\end{equation}
So the stationarity condition gives
\begin{equation} \label{eq:4.1-05}
\int d^2 a_2 \delta(V') = \int d^2 a_2 M_S^2 \frac{\lvert \det V'' \rvert}{\lvert J \rvert} \delta^2 (a_2 - a_{2(0)})
                     \sim F^2 M_S^4 (\frac{1}{M_K^4} + \frac{1}{M_P^4}) \ .
\end{equation}
And the metastability condition gives
\begin{equation} \label{eq:4.1-06}
\int d^2 a_3 \Theta(V'' > 0) \sim \int d^2 a_3 \Theta(a_3 \lesssim F M_S^2 (\frac{1}{M_K^2} + \frac{1}{M_P^2}))
                             \sim F^2 M_S^4 (\frac{1}{M_K^4} + \frac{1}{M_P^4}) \ .
\end{equation}
From \eqref{eq:4.1-03} we know the integration region of $a_0$ is restricted in
\begin{equation} \label{eq:4.1-07}
\lvert a_0 \rvert \in (\frac{M_P}{\sqrt{3} M_S} \sqrt{F^2 - \frac{\Lambda_0}{M_S^4}}, \frac{M_P}{\sqrt{3} M_S} F)
              \approx (\frac{M_P}{\sqrt{3} M_S} (F - \frac{\Lambda_0}{2 F M_S^4}), \frac{M_P}{\sqrt{3} M_S} F)
\end{equation}
which is a similar ring-shaped area as shown in figure \ref{fg:1}. So the small cosmological constant condition gives
\begin{equation} \label{eq:4.1-08}
\int d^2 a_0 \Theta(0 <  V  < \Lambda_0) \sim 2 \pi \frac{F M_P}{\sqrt{3} M_S} \cdot \frac{\Lambda_0 M_P}{2 \sqrt{3} F M_S^5}
                                         \sim \frac{\Lambda_0 M_P^2}{M_S^6} \ .
\end{equation}

Combining \eqref{eq:3.0-01}\eqref{eq:4.1-05}\eqref{eq:4.1-06}\eqref{eq:4.1-08} we get the number of SUSY-breaking vacua for the general SUGRA case:
\begin{equation} \label{eq:4.1-09}
\begin{split}
N(F < F_0, 0 <  V  < \Lambda_0) &\sim \int d^8 a_0 \dotsm a_3 \Theta(F < F_0) \delta(V') \Theta(V'' > 0) \Theta (0 <  V  < \Lambda_0)\\
                                &\sim \int_0^{F_0} F \cdot F^2 M_S^4 (\frac{1}{M_K^4} + \frac{1}{M_P^4}) \cdot F^2 M_S^4 (\frac{1}{M_K^4} + \frac{1}{M_P^4}) \cdot \frac{\Lambda_0 M_P^2}{M_S^6} d F\\
                                &\sim F_0^6 \Lambda_0 M_S^2 M_P^2 (\frac{1}{M_K^8} + \frac{1}{M_P^8}) \ .
\end{split}
\end{equation}
In most realistic models $M_K \lesssim M_P$, so we have
\begin{equation} \label{eq:4.1-10}
N(F < F_0, 0 <  V  < \Lambda_0) \sim F_0^6 \frac{\Lambda_0 M_S^2 M_P^2}{M_K^8} \ .
\end{equation}
Comparing to the result \eqref{eq:3.3-07} before including mass scales, the vacuum distribution has the same form respecting to $F$ and $\Lambda$. The power of $F_0$ indicates that the distribution is peaked at the highest SUSY breaking scale. Most states live around the cut-off $F \sim F_0$. The distribution respecting to the cosmological constant is flat. But now we have the extra factor from including mass scales. With the mass hierarchy presumed, we have made our analysis procedure valid up to $F_0 \sim 1$. The distribution is reliable all the way up to the cut-off. But the tail of the distribution after the cutoff could still be complicated.

\subsection{The Global SUSY and canonical K\"ahler potential limit}

If we take the limit $M_P \to \infty$ of \eqref{eq:4.1-09}, the number of vacua goes to $\infty$. This is because there is a factor coming from the small cosmological constant condition which set the value of $a_0$ having an allowed area proportional to $M_P^2$. As we have discussed before, the small cosmological constant condition should be dropped off in the global SUSY case. Most analysis goes through similarly to the SUGRA case. Combining \eqref{eq:3.0-01}\eqref{eq:4.1-05}\eqref{eq:4.1-06} and taking the limit $M_P \to \infty$, we get the number of SUSY-breaking vacua for the global SUSY case with a general K\"ahler potential:
\begin{equation} \label{eq:4.2-01}
N(F < F_0) \sim \int d^6 a_1 \dotsm a_3 \Theta(F < F_0) \delta(V') \Theta(V'' > 0)
           \sim \int_0^{F_0} F \cdot F^2 \frac{M_S^4}{M_K^4} \cdot F^2 \frac{M_S^4}{M_K^4} d F
           \sim F_0^6 \frac{M_S^8}{M_K^8} \ .
\end{equation}
Comparing to the result \eqref{eq:3.2-03} before including mass scales, the vacuum distribution has the same form respecting to $F$. But now we have the extra factor from including mass scales.

To see the vacuum distribution with a canonical K\"ahler potential, we take the limit $M_K \to \infty$, \eqref{eq:4.1-09} becomes
\begin{equation}
N(F < F_0, 0 <  V  < \Lambda_0) \sim F_0^6 \frac{\Lambda_0 M_S^2}{M_P^6} \ .
\end{equation}
The distribution respecting to $F$ and $\Lambda$ has the same form as \eqref{eq:4.1-09}. This explains what we have seen in the SUGRA case before including mass scales where we get the same distribution with an either canonical or general K\"ahler potential.

If we take the limit $M_K \to \infty$ of \eqref{eq:4.2-01}, the number of vacua goes to $0$. This shows that metastable SUSY breaking is rare in generic global SUSY models with a canonical K\"ahler potential unless one introduces R-symmetries, as we have discussed before.

\subsection{Comparing to SUSY vacuum distributions}

Since we only assumed a uniform parameter distribution in general cases, the vacuum distributions what we have got before have only relative meaning. Even though without knowing the the detailed form of $d \mu(a_n, c_{nm}, z)$ from the microscopic landscape, the factors we have got from including mass scales enable us to do a general comparison between the distributions of non-SUSY and SUSY vacua, which is an interesting topic to study in the landscape.

We start again from \eqref{eq:4.1-01}--\eqref{eq:4.1-02}. For SUSY vacua, the F-term vanishes, so $a_1 = 0$. Then we always have $V \le 0$. In the SUGRA case it is still worth to count the vacua with $- \Lambda_0 < V < 0$ and expect that other corrections of the same order could set the cosmological constant to a positive value. To get such a vacuum requires
\begin{equation} \label{eq:4.3-01}
0 < \lvert a_0 \rvert < \sqrt{\frac{\Lambda_0}{3}}\frac{M_P}{M_S^3} \ .
\end{equation}
$\partial V$ always vanish for $a_1 = 0$, so the stationarity condition gives no extra constraint. The elements of the mass matrix become
\begin{align}
\partial^2 V &= - a_0^* a_2 \frac{M_S^4}{M_P^2} \sim a_2 \frac{\sqrt{\Lambda_0} M_S}{M_P} \ ,\\
\bar{\partial} \partial V &= a_2^* a_2 M_S^2 - 2 a_0^* a_0 \frac{M_S^6}{M_P^4} \sim a_2^* a_2 M_S^2 \ .
\end{align}
The dominant term of $\bar{\partial} \partial V$ has a positive definite form and $\lvert \partial^2 V \rvert \ll \bar{\partial} \partial V$. $\det V'' > 0$ is always satified except in an exceptionally small parameter space region where $a_2$ is of order $\sqrt{\Lambda_0} / (M_S M_P)$. So the metastability condition also gives no extra constraint.

There is no constraint on $a_n$ for $n > 1$. They can take any value up to order $1$. It is $a_1 = 0$ that satisfies the stationarity condition. So we treat $\partial V$ as a function of $a_1$ and $a_1^*$. Then we have
\begin{equation}
\lvert J \rvert = \left \lvert \frac{\partial (\partial V, \bar{\partial} V)}{\partial (a_1^*, a_1)} \right \rvert
= a_2^* a_2 M_S^6 - 4 a_0^* a_0 \frac{M_S^{10}}{M_P^4} \sim M_S^6 \ .
\end{equation}
From the estimation of elements of $V''$, we have
\begin{equation}
\lvert \det V'' \rvert \sim (a_2^* a_2)^2 M_S^4 - a_2^* a_2 \frac{\Lambda_0 M_S^2}{M_P^2} \sim M_S^4 \ .
\end{equation}
So the stationarity condition gives
\begin{equation} \label{eq:4.3-02}
\int d^2 a_1 \delta(V', F = 0) = \int d^2 a_1 M_S^2 \frac{\lvert \det V'' \rvert}{\lvert J \rvert} \delta^2 (a_1) \sim 1 \ .
\end{equation}
The metastability condition gives only an order $1$ factor. From \eqref{eq:4.3-01} we know the small cosmological constant condition gives
\begin{equation} \label{eq:4.3-03}
\int d^2 a_0 \Theta(- \Lambda_0 <  V  < 0) \sim \frac{\Lambda_0 M_P^2}{M_S^6} \ .
\end{equation}

Combining \eqref{eq:4.3-02}\eqref{eq:4.3-03} we get the number of SUSY vacua for the SUGRA case:
\begin{equation}
N(F = 0, - \Lambda_0 <  V  < 0) \sim \int d^8 a_0 \dotsm a_3 \delta(V', F = 0) \Theta(V'' > 0) \Theta (0 <  V  < \Lambda_0) \sim \frac{\Lambda_0 M_P^2}{M_S^6} \ .
\end{equation}
For the global SUSY case, most analysis is similar to the SUGRA case. Dropping off the small cosmological constant condition and taking the limit $M_P \to \infty$, we get the number of SUSY vacua:
\begin{equation}
N(F = 0) \sim \int d^6 a_1 \dotsm a_3 \delta(V', F = 0) \Theta(V'' > 0) \sim 1 \ .
\end{equation}
Now we can compare these results to \eqref{eq:4.1-10} and \eqref{eq:4.2-01}. Setting $F_0 \sim 1$, we have
\begin{gather}
\frac{N(F \sim 1)}{N(F = 0)} \sim \frac{M_S^8}{M_K^8} \quad
\text{for SUSY},\\
\frac{N(F \sim 1, 0 <  V  < \Lambda_0)}{N(F = 0, - \Lambda_0 <  V  < 0)} \sim \frac{M_S^8}{M_K^8} \quad
\text{for SUGRA}.
\end{gather}
In both SUSY and SUGRA cases, if all mass scales are set to be $1$, the numbers of SUSY and non-SUSY vacua are comparable when the cutoff $F_0$ is set to be of order $1$. If we assume $M_S \ll M_K$ for low energy model building, non-SUSY vacua are always rare compared to SUSY vacua by the factor of $M_S^8 / M_K^8$.

\section{Parameter tuning}

In previous sections we have stated many times that the result indicates the rareness of SUSY breaking. At the end of the last section we even get the rareness quantitatively. These results come from the assumption that Lagrangian parameters have a smooth distribution and can be approximated by a uniform distribution. Such an assumption can not go beyond the tree level SUSY breaking branch of the landscape. In the intermediate or low energy SUSY breaking branch where SUSY is broken dynamically, there are singularities in the parameter distribution, and the vacuum distribution regarding to $F$ is very different than the tree level SUSY breaking branch \cite{Banks:2003es, Dine:2004is, Dine:2005iw}. Besides the counting of metastable vacua, non-perturbative instability may furthermore alter the distribution \cite{Dine:2007er, Dine:2008jx, Dine:2009tv, Narayan:2010em}. Cosmological settings such as chaotic inflation also realize vacua with different probabilities \cite{Linde:1983gd, Winitzki:2006rn, Guth:2007ng}. For low energy model building, it is more important to know which parameter region leads to phenomenologically acceptable models rather than having the vacuum distribution. One can ask how much tuning is needed to get a desired vacuum with appropriate $F$ and $\Lambda$, and leave the question of the origin of the tuning to the future study of fundamental theories. Although it is quite a trivial routine to work out the allowed parameter region from a specific model, our EFT method provides a general view to the amount of tuning. In previous sections, constraining Lagrangian parameters is a procedure to calculate the vacuum distribution. But the constrained region of $a_n$'s is just what we concern here.

We assume $M_S \ll M_K < M_P$ and $F \sim 1$. In the SUGRA case, \eqref{eq:4.1-07} shows $a_0$ is restricted in a ring-shaped area as shown in figure \ref{fg:1}, with its magnitude around
\begin{equation}
a_0 \sim \frac{M_P}{\sqrt{3} M_S} F \sim \frac{M_P}{M_S} \ .
\end{equation}
The large value of $a_0$ seems unnatural, but is required to cancel the contribution from $D W$ if there is no other contribution to the vacuum energy. SUSY breaking gives the gravitino mass
\begin{equation}
m_{3/2} = \frac{W}{M_P^2} = a_0 \frac{M_S^3}{M_P^2} \sim \frac{M_S^2}{M_P}
\end{equation}
which means that SUSY breaking occurs at an intermediate scale between $m_{3/2}$ and $M_P$. This is actually a feature of phenomenological SUGRA models, i.e.\ gravity mediation. As a comparison, other scenarios such as gauge mediation favor low scale SUSY breaking.

In the previous treatment, $a_n$'s are scaled to dimensionless by powers of $M_S$. We have assumed that typical values of $a_n$'s are of order $1$ if they are generated by the same dynamics which breaks SUSY. But $a_0$ does not contribute to SUSY breaking except for adjusting the cosmological constant\footnote{Even $a_0$ appears in $D W$ in the SUGRA formula, it is suppressed by powers of $M_S / M_P$ or $M_K / M_P$.}. It is reasonable to think that $a_0$ gets contribution from processes other than the SUSY breaking dynamics and not related to $M_S$. Alternatively, one can introduce another hidden sector which dynamically generates $a_0$ with its magnitude to the correct order, as what has been done in retrofitting models \cite{Dine:2006gm}. However, the narrow thickness of the ring-shaped area as shown in figure \ref{fg:1} suggests that an order $\Lambda_0 / M_S^4$ tuning is necessary for a small cosmological constant.

Now we focus on constraints on other parameters which are common in both global SUSY and SUGRA cases. We have seen $\lvert a_1 \rvert = F$ which is just the SUSY breaking field strength. The value of $a_2$ is determined from $a_1$ by a delta function. This is just the requirement of a stationary vacuum and should not be viewed as tuning. In fact, when we do the translation $(z-z_0) \to z$ to set the vacuum at $z=0$, the value of $z_0$ occupies two dimensions of the parameter space which compensate the lost two dimensions from fixing $a_2$. The only tuning comes from the metastability condition. The constraint \eqref{eq:4.1-04} implies
\begin{equation} \label{eq:5.0-01}
a_3 \lesssim F \frac{M_S^2}{M_K^2} \sim \frac{M_S^2}{M_K^2} \ .
\end{equation}
Since $a_3$ takes a complex value, the needed amount of tuning for metastable SUSY breaking is of order $M_S^4 / M_K^4$. Conversely, the metastability condition for SUSY vacua is always satisfied in the global SUSY case and only excludes exceptionally small parameter region in the SUGRA case. No tuning is needed to get a SUSY vacuum. Note that we have done several field redefinitions to simplify the form of the potential. The relation between $a_n$'s and parameters of the original model could be complicated. So the amount of tuning may be spreaded in several of the original parameters when we try to build a realistic model.

Our analysis up to now is based on tree level calculation. One may argue that loop corrections could have extra contribution to the mass of $z$, thus relaxing the tuning of \eqref{eq:5.0-01}. One example comes from direct mediation where the SUSY breaking field $z$ couples to some messenger fields $\phi_i$ in the same hidden sector. The pseudomodulus gets a Coleman-Weinberg mass of order $M_S^2 / M_{\phi}$ where $M_{\phi}$ is the typical messenger mass \cite{Coleman:1973jx}. The Coleman-Weinberg formula is only valid when $M_{\phi} \gg M_S$. So the contribution to the mass of $z$ is much smaller than $M_S$. One still needs some tuning for $a_3$ to keep the vacuum metastable. In fact, one can integrate out heavy fields when $M_{\phi} \gg M_S$ and get non-minimal corrections to the K\"ahler potential. Then the $M_K$ can be substituted by $M_{\phi}$ in the tuning of \eqref{eq:5.0-01}.

Although much tuning is needed when $M_S \ll M_K$, there is always some allowed parameter region as long as non-minimal corrections of the K\"ahler potential do not vanish. Precisely speaking, if we follow the procedure how we get \eqref{eq:4.1-04}, the leading order contribution to \eqref{eq:5.0-01} is from $c_{12}^2$ and $c_{22}$. If one can freely choose the form of the superpotential and the K\"ahler potential, non-vanishing $c_{12}$ or $c_{22}$ generally allows one to build models with metastable SUSY breaking vacua in some parameter region. This feature has been observed previously while building models locally equivalent to the Polonyi model \cite{Kitano:2008tm}. Here our result shows that the argument is true for most general F-term SUSY breaking models in both global SUSY and SUGRA cases. The size of the allowed parameter region is always proportional to $M_S^4 / M_K^4$.

We have not introduced any symmetry in our analysis. In SUSY breaking model building, R-symmetries are often applied because of their connection to SUSY breaking global vacua \cite{Nelson:1993nf}. Usually the pseudomodulus $z$ has R-Charge $2$. $z$ can only appear linearly in the superpotential because of the R-symmetry\footnote{However with coupling to other fields, a term like $z z \phi_i$ is not allowed for metastability reasons even if they may be allowed by the R-symmtery. The lowest allowed term non-linear in $z$ is $z z \phi_i \phi_j$ which only appears in a non-renormalizable superpotential.}. Although only part of the parameter space can make a metastable SUSY breaking vacuum, the reduction is usually of order $1$ and no tuning is required. Once the R-symmetry is broken, terms like $a_3 z^3$ may appear in the superpotential and destroy the metastability. So the tuning of \eqref{eq:5.0-01} also characterizes the allowed amount of R-symmetry breaking which can be introduced into R-symmetric models. We see again the importance of approximate R-symmetries in metastable SUSY breaking model building, as discussed in literatures for different reasons \cite{Intriligator:2007py, Abe:2007ax}.

Finally, we would like to comment on extending the one-field approximation to multi-field cases. We only studied the behaviour of the potential along the pseudomodulus direction in our current analysis. Other fields, if they are heavy than the SUSY breaking scale, could be viewed as appearing in the non-minimal corrections of the K\"ahler potential after being integrated out. If more than one fields are light, one can still identify the pseudomodulus-goldstino along the F-term direction. The same tuning of \eqref{eq:5.0-01} is needed following the same analysis of the one-field case. In addition, stabilizing other field directions may further reduce the allowed parameter region, either by a factor of order $1$ or more tuning. Having more than one light field seems statistically rare, but may be preferred by symmetry consideration or for solving phenomenology problems \cite{Cheung:2010mc, Argurio:2011hs, Cheung:2011jq}.

\section*{Acknowledgement}

The author would like to thank Michael Dine, Sandip Trivedi and Satoshi Yamaguchi for helpful discussions.

\end{document}